\documentstyle[amssymb,12pt]{article}

\input{tcilatex}

\begin{document}

\title{Monte Carlo simulation of the transmission of measles: Beyond the mass
action principle}
\author{Nouredine Zekri$^{1,2 \thanks{%
Email zekri@mail.univ-usto.dz}}$ and Jean Pierre Clerc$^{2 \thanks{
Email clair@iusti.univ-mrs.fr}}$ \\
{\small $^{1}$U.S.T.O., D\'{e}partement de Physique, L.E.P.M., B.P.1505 El
M'Naouar, Oran, Alg\'{e}rie. } \\
{\small $^{2}$I.U.S.T.I., Technop\^{o}le Ch\^{a}teau Gombert, Universit\'{e}
de Provence, Marseille, France. }}
\maketitle

\begin{abstract}
\hspace{0.33in} We present a Monte Carlo simulation of the transmission of
measles within a population sample during its growing and equilibrium states
by introducing two different vaccination schedules of one and two doses. We
study the effects of the contact rate per unit time $\xi $ as well as the
initial conditions on the persistence of the disease. We found a weak effect
of the initial conditions while the disease persists when $\xi $ lies in the
range $1/L$-$10/L$ ($L$ being the latent period). Further comparison with
existing data, prediction of future epidemics and other estimations of the
vaccination efficiency are provided.

\hspace{0.33in} Finally, we compare our approach to the models using the
mass action principle in the first and another epidemic region and found the
incidence independent of the number of susceptibles after the epidemic peak
while it strongly fluctuates in its growing region. This method can be
easily applied to other human, animals and vegetable diseases and includes
more complicated parameters.
\end{abstract}


\noindent PACS Number(s): 02.50.Ng, 87.53.Wz, 02.50.-r, 89.65.-s

\newpage

\section{Introduction}

\hspace{0.33in} The mathematical investigation of the disease transmission
was initiated by Daniel Bernoulli three centuries ago \cite{Berno}, but this
field started to grow considerably only at the begining of the twentieth
century when Hamer, in his attempt to understand the recurrence of measles
epidemics assumed that the incidence (rate of new cases) depends on the
product of the densities of susceptibles and infectives \cite{Hamer}. This
assumption is now the basis of the modern mathematical epidemiology and is
known as the {\it mass action principle}. Epidemiological models using this
principle are widely reviewed in the literature [3-5]. Basically, they
formulate the flow patterns between three classes of population: the
Susceptibles ($S$), the Infectives ($I$) and the Recovered ($R$) which are
immune either by a vaccination or by the disease itself. Some more
complicated models increase the number of classes to five by including the
class of passively immune individuals due to maternal antibodies ($M$) and
that of Exposed ones ($E$) during the latent period \cite{Ander, Hethc}. The
mathematical formulation of a classical $SIR$ model yields a set of coupled
first order differential equations

\begin{eqnarray}
\frac{dI}{dt} &=&\beta I\frac{S}{N}-\gamma I-\mu I \\
\frac{dS}{dt} &=&-\beta I\frac{S}{N}+m-\mu S \\
\frac{dR}{dt} &=&\gamma I-\mu R
\end{eqnarray}

\noindent Here $N$ denotes the total population number, $\beta $ the
transmission rate, $\gamma $ the recovery rate, $\mu $ the death rate and $m$
the number of births per unit time. When the population reaches its
equilibrium, the rate of births becomes equal to that of deaths ($m=\mu N$).
The transmission rate depends in general both on the infective and infected
ages and becomes a continuous matrix with elements $\beta _{aa^{\prime }}$
describing the probability by unit time to infect a susceptible of age $a$
by an infective of age $a^{\prime }$. It is in practice impossible to solve
the above equations with such a continuous matrix which is reduced in
general to a $5x5$ matrix within cohorts of age called $WAIFW$ \cite{Ander,
Hethc}. This rate depends also on genetic and spatial heterogeneities. In
the case of measles there is no genetic heterogeneity but the spatial
dependence has been shown by some geometrical models like Small World
networks \cite{Watts} and cellular automata \cite{Bocca} to affect
sensitively the dynamics of epidemics. Some empirical models were proposed
within the framework of the $SIR$ equations \cite{Ander, Hethc} to take into
account the spatial heterogeneity either by including an $N$ dependence of $%
\beta $ with an exponent $\upsilon $ or proposing more complicated
algebraically expressions to $\beta $, but this rate depends in fact on a
complicated distribution of the number of acquaintances between individuals 
\cite{Watts, Zekri}. Models based on the previous equations are widely used
both to predict epidemics \cite{Rober}, or to optimize the vaccination
schedule \cite{Levy} or even to find the critical coverage to eradicate the
disease \cite{Ander}.

\hspace{0.33in} However, nobody checked the validity of the mass action
principle. In particular, since only infectives and susceptibles contribute
to the incidence, the $N$ dependence of this principle is questionable at a
first look. In Eqs.(1-3) it is easier to handle only the of $S$, but this
could lead to an under estimation of the data as found in New Zealand \cite
{Rober} where some epidemics were observed before the predicted dates.
Furthermore, the rate $\beta $ (which is fixed in the above equations)
should change asymptotically depending on whether the incidence is much
larger or much smaller than the number of receptives $S$.\ 

\hspace{0.33in} On the other hand, the estimation of the vaccination
efficiency should take into account the distribution of susceptibles within
the cohorts of vaccination age. For example, suppose in a given population
all susceptibles are within ages below that at which the vaccination holds,
it will not be efficient. The numbers of susceptibles by cohorts of ages are
impossible to measure for the whole population because they need an
extensive serological investigation which is very expensive. Therefore, it
seems necessary to simulate by a new approach the dynamics of the infection
in order to examine the temporal behavior of the disease and also to check
the mass action principle for different conditions.

\hspace{0.33in} This is the aim of the present paper where we use a Monte
Carlo simulation of measles propagation in a population for a period of $250$
years (from $1850$ to $2100$) in order to take into account the growing
periods in the steady state and the equilibrium ($N$ constant). We
investigate the effect of the initial conditions and the contact rate on the
temporal dependence of the infection as well as the vaccination efficiency
(we have introduced two different vaccination schedules).\ We compare our
numerical data to the existing one in Oran (Algeria) and our incidence to
that of Eqs.(1,2) based on the mass action principle.

\section{Description of the Method}

\hspace{0.33in} The present algorithm is inspired from that used in particle
physics (GEANT) where particles are followed within different detectors to
estimate their geometrical acceptance \cite{Geant}. We start increasing time
by steps of one day from the date $0$ (i.e. $1/1/1850$) towards $250$ years.
In each step we generate $m$ new births and attribute to them maternal
antibodies with random live times following an exponentially decreasing
distribution vanishing after $15$ months \cite{Ander, cdc}, the rate of this
decrease is determined by imposing $20\%$ of children remaining naturally
immune at $9$ months age. When loosing their natural antibodies they change
to the susceptible class and the vector element $S(a)$ of age $a$ (in days)
is incremented. In the same step we generate $m$ times the death age $%
a^{\prime }$, from a distribution centered at the live expectancy and remove
the corresponding susceptibles from the $S$ vector. We consider $m$ constant
($m=100$) in this work, leading in the absence of infectives to a linear
growth with time of the number susceptibles ($S=mT$). We assume also a
delta-peak distribution of deaths at $100$ years in order to ensure a
constant total population at equilibrium and show the dependence on $N$ of
the incidence in this case. This choice will affect only the equilibrium
date since the infection holds mainly below $20$ years age. After a time $T$%
, we introduce an external infective individual (coming from another sample)
assumed at the end of its latent period which is removed after its infection
period ($7$ days). The time $T$ allows us to adjust the number of initial
susceptibles before the infection. This infective {\it attacks} $\xi $
susceptibles per day ($\xi $ is the contact rate per unit time and per
infective and corresponds to $\beta S/N$ in the above equations). The new
infected persons are removed from the corresponding elements of the vector $%
S $ and increment the number of infected $I$, and after the latent period ($%
L=7 $) they become infective and increment the number of infectives $I_{1}$
and remove them from $I$, before being removed from $I_{1}$ at the end of
their infection period so that the recovery period $\gamma ^{-1}$ is $14$
days. During their infection period each new infective {\it attacks} an
average of $\xi $ susceptibles per day and so on. The age of the
susceptibles to be infected is generated randomly with a probability
distribution

\begin{equation}
P(a_{i})=\left\{ 
\begin{array}{l}
25\%\text{ ; if }a_{i}\leqslant 5\text{ years} \\ 
45\%\text{ ; if }5<a_{i}\leqslant 10\text{ years} \\ 
20\%\text{ ; if }10<a_{i}\leqslant 15\text{ years} \\ 
9\%\text{ ; if }15<a_{i}\leqslant 20\text{ years} \\ 
1\%\text{ ; if }a_{i}>20\text{ years}
\end{array}
\text{ }\right\}
\end{equation}

\noindent We vary this age within $\pm 2$ years until finding a susceptible
having an age in this range. If not found, we do not increment $I$. We
neglect the spatial heterogeneity by considering an average contact rate (it
is possible to use a distribution), and assume the incidence independent of
the age of infectives which corresponds to the case $WAIFW$ $3$ (it is
possible to include the age of infectives by using in this algorithm $I$ and 
$I_{1}$ as vectors). The probability distribution $P(a)$ was choosed from
the average distribution of infected persons within the same cohort of age
in developing countries where the infection seems to have an average age in
the scholar ages \cite{Mlean}. In developed countries this average age is
slightly higher.

\hspace{0.33in} We introduce two vaccination programmes in periods
corresponding to the case of Algeria. The first vaccination schedule in $%
1970 $ provides one dose at $9$ months and the second one in $1995$
providing two doses at $9$ months and $2$ years (other ages of the second
dose administration are discussed below to compare with the data of
Algeria). Note here that the first dose fails for all persons losing their
maternal antibodies after $9$ months. The coverage is fixed at $80\%$ in the
rest of the paper. These parameters simulate the average situation in the
countries administrating a double dose vaccination schedule. The quantities $%
S$, $I$ and $I_{1}$ are recorded by steps of $14$ days while $%
I^{-1}S^{-1}dI/dS$ (the incidence per infective and per infected) is
recorded by steps of one day to compare our incidence with Eqs.(1,2). In
order to show the persistence of the disease within the whole period of the
simulation, we need a sufficiently high contact rate between susceptibles
enabling the first epidemic peak. In a recent paper, we found within the
framework of small world network the number of connections between
susceptibles for a childhood disease randomly distributed between $1$ and $6$
at the threshold concentration of susceptibles leading to an epidemic state
(see fig.3b of \cite{Zekri}). Therefore, since the infection period is $7$
days, an epidemic situation holds if we choose $\xi =0.5$ corresponding to
an average number of acquaintance of about $3$. The value of $\xi $ is fixed
for the rest of the paper except when it is varied.\ This algorithm takes an
average computation time of about $5$ hours in Pentium III 600Mhz personal
computers. It is obviously possible to introduce in this algorithm
complicated distributions of $\xi $, birth and death rates, vaccination age
and latent period, but we fix them in this paper.

\section{Results and discussion}

\hspace{0.33in} In figures 1, we show the temporal dependence of the number
of infected and susceptible persons for two different situations of the
initial date of infection: $T=2$ days and $20$ years, to compare the cases
of low and high density of susceptibles with respect to the total population
number. For the demographic situation described in this work ($m$ constant),
the first case corresponds to only $100$ susceptibles while in the second
one we have $730000$ susceptibles in the sample. After the first strong
epidemic peak whose amplitude depends on the number of susceptibles, the
number of infected (and susceptibles), which oscillates as a function of
time in the steady state, shows a similar amplitude for the two initial
conditions with a slight variation of the inter-epidemic period (due to the
nonlinear aspect of the infection process), which can influence the measure
of the number of predicted cases. On the other hand, since measles appeared
before the 19th century the first case ($T=2$ days) appears more realistic.
The persistence of the disease concerns the whole period of the simulation
(in Fig.1a), although we introduced a two dose vaccination schedule. The one
and two doses schedules seem to decrease the average numbers of susceptibles
and infected peaks by about $60\%$ and $90\%$ respectively. A simple
estimation taking into account the coverage and the vaccination failure due
to its annihilation with maternal antibodies yields rates slightly higher ($%
64\%$ for the one dose and $94\%$ for two doses). In fact this estimation
assumes that the elements of the vector $S$ are filled ($100$ susceptibles)
at the age of the vaccination which is not the case here.

\hspace{0.33in} The effect of the contact rate $\xi $ in the case $T=2$ days
is shown in figures 2. This parameter seems to affect significantly both the
incidence and the inter-epidemic period. The amplitudes of the infection
peaks (and the average number of susceptibles) decrease as $\xi $ increases.
The first epidemic peak is more stronger than the other ones and occurs
after a period decreasing as $\xi $ increases (the same behavior holds for
the inter-epidemic period). Indeed, initially the number of susceptibles is
much smaller than the incidence which decreases the effective contact rate $%
\xi _{eff}=P(a)\xi $ below the recovery rate ($\xi _{eff}<\gamma $), while
the number of susceptibles grows linearly with time in this period. As a
consequence, the propagation remains endemic until the time $T_{1}$ (at
which the first epidemic peak occurs) satisfying the following equation

\begin{equation}
\xi P(T_{1})-2\gamma =0
\end{equation}

\noindent the factor $2$ means that the first term concerns the infectives $%
I_{1}$ while the second one is the rate of all infected $I$ which are twice $%
I_{1}$, it corresponds to $1/L\gamma $ in a general disease. If $T_{1}$ is
within the same class, it behaves as $\xi ^{-1}$ with a threshold $\xi
_{\min }=2\gamma $. The number of susceptibles (Fig.2b) reaches its maximum
when the total incidence corresponds to the birth rate ($\xi _{eff}PI_{1}=m$%
), while after the epidemic peaks $\xi _{eff}=\xi _{\min }$ and $S$ reaches
its minimum.\ The inter-epidemic period seems to decrease slightly during
the vaccination periods but the amplitude of the peaks depends on the number
of susceptibles at the age of the vaccination which decreases as $\xi $
increases. This can be seen in Fig.2a for $\xi =0.3$ where both the first
and second vaccination schedules decrease sensitively the epidemic peaks,
while if $\xi $ increases this effect becomes weaker up to a critical
contact rate where all susceptibles have smaller ages than the vaccination
one. This is shown for $\xi =2$ where the first epidemic peak is as strong
as all susceptibles are infected and the effective value of this rate is
sufficiently large to infect all new susceptibles (see figure 2a), so that $%
S=0$ and the number of infected remains constant $I=m/\gamma $ during the
whole period.\ This is the case if the minimum value of $\xi _{eff}$ is
larger than $2\gamma $. Indeed, since the age of infection varies within a
range of $\pm 2$ years, even if there are susceptibles having only $1$ day,
from Eq.(4) they are infected with a probability of $10\%$ and the
corresponding effective contact rate is $\xi /10$.\ Consequently the
critical contact rate above which all the new susceptibles are infected is ($%
\xi _{c}=20\gamma =10/L$). Note in this case that both vaccination schedules
do not affect the number of infected nor susceptibles since all the infected
persons are within their first year age.\ Therefore, the range of the
contact rate necessary to ensure a persistent disease propagation is

\begin{equation}
1/L<\xi <10/L
\end{equation}

\noindent In the asymptotic limit of vanishing $L$ the disease does not
persists, while if $L$ is very large any small contact rate makes the
disease persistent.

\hspace{0.33in} Now let us compare our results with the existing data of
Oran (Algeria) in the period $1994-1997$ \cite{Benth}. We use the second
dose at $6$ years which is the case of this country. A good agreement is
shown for the period $1994-1996$, while in $1997$ the number of cases is
significantly different (as shown in table I).\ We think that the contact
rate $\xi $ used here simulates a real one but the birth rate could change
leading to an increase of the inter-epidemic period.\ Indeed, the birth
rate, has significantly decreased in this country at the end of the
eighteens due to different social events.\ The vaccinated children in $1996$
were born in $1990$ and their rate decreased implying an increase of the
inter-epidemic period which was initially about $2$ years. Therefore, a more
accurate simulation should take into account the real variation of the
demographic situation.

\hspace{0.33in} The effect of the age of the second dose on the incidence of
the virus\ is shown in figure 3, where we compare in the period $1990-2020$
the evolution of the number of infected for two different ages ($2$ and $6$
years). The vaccination at $2$ years decreases significantly the infected
peaks in comparison with that at $6$ years, while it increases the
inter-epidemic period. From this figure, we deduce a reduction of the
predicted numbers of infected of about $87\%$, $77\%$, $78\%$ and $79\%$ for
the periods between $1995$ and $2000$, $2005$, $2010$, and $2020$,
respectively. We expect from $1995$ to $2020$ about $12800$ cases if the
vaccination is at $6$ years and about $2740$ cases at $2$ years. We expect
also from the above comparison with the data in Oran an epidemic peak in
this city in $2002$ if the birth rate does not change.

\hspace{0.33in} Let us now compare our results to those using the mass
action principle in Eqs. (1,2). Already from Figs.1 and 2, we deduce that
both the incidence and the serological situation (number of susceptibles)
does not depend on the total population number since the oscillations remain
with the same amplitude and period (inter-epidemic period) both in the
growing period (before $1950$) and the equilibrium one (after this date $N$
is constant). In figure 4, we show the variation of $I^{-1}S^{-1}dI/dt$ with
the number of susceptibles in two different regions : the two sides of the
first epidemic peak (Fig.4a,b) and those of another one at equilibrium
(Fig.4c,d). We see in the regions where $I$ increases (Figs. 4a, c) strong
fluctuations of this quantity with $S$ with a power law variation in a part
of Fig.4c, in the case of a small number of susceptibles (due to an endemic
peak). In the regions of decreasing $I$ (Figs. 4b, d) this quantity
decreases as a power-law with $S$ with an exponent close to unity,
indicating the independence of the incidence in Eqs. (1 and 2) on the number
of susceptibles. This discrepancy with the mass action principle is due to
the saturation effects shown in our approach based on the contact rate $\xi $%
. Indeed, the behavior in the decreasing regions (after the peak) is due to
the fact that the number of susceptibles is much smaller than the incidence $%
\xi I_{1}$ ($I_{1}$ is large in this region) which varies in this case
independently of $S$ (the decrease is mainly governed by the recovery rate $%
\gamma $). While in the increasing regions, the fluctuations are due to the
stochastic behavior of the distribution $P(a)$ in Eq.(4). Therefore, the
expression of the incidence in the mass action principle should take into
account these asymptotic situations based on the contact rate for a
realistic prediction of the disease propagation.

\section{Conclusion}

\hspace{0.33in} We presented a Monte-Carlo simulation of measles
transmission in a population having a constant birth rate within $250$
years. In this simulation we introduced realistic parameters like
distribution of the live time of the maternal antibodies, the latent and
infection  periods and two different vaccination schedules. We fixed the
live expectancy at $100$ years to ensure a constant population number.
Although, this value is high and not realistic but it does not change the
general behavior of our results. We found the disease persistent for a
contact rate $\xi $ between $1/L$ and $10/L$ with inter-epidemic periods
depending on $\xi $ and $m$, while the disease propagation is only slightly
affected by the initial conditions although we expected a non linear
behavior predicting an unstable evolution. This investigation allowed us to
predict both the efficiency of the vaccination and future epidemic peaks. We
compared our simulation results with existing data (from $1994$ to $1997$)
in the city of Oran (Algeria) which has a similar demographic situation as
well as approximately identical periods of the vaccination as used in this
paper. We found a good agreement for the $3$ first years while the number of
simulated cases is higher for the fourth one, probably due to the variation
of the birth rate at the beginning of the last decade in this country. We
showed also that a second dose vaccination at $2$ years reduces the number
of cases by more than $75\%$ within the next two decades in comparison with
a vaccination at $6$ years (occurring in Algeria). This suggests that $75\%$
of the children between $2$ and $6$ years are naturally infected. Finally we
compared our results to the incidence suggested by the mass action principle
and found a discrepancy due to the asymptotic behavior of the incidence when
the number of susceptibles is small compared to the incidence.

\hspace{0.33in} Therefore, this algorithm seems to be more realistic and can
be easily extend it to other human, animal and vegetable diseases only by
introducing the corresponding parameters and distributions. Since, the
spatial heterogeneity was characterized by a distribution of connections in
a recent work on the small world network \cite{Zekri} and the disease
propagation depends mainly on the connections, we can introduce this
heterogeneity only by including the right distribution. It is also possible
to combine this algorithm to the small world network to take into account
the spatial heterogeneity. \bigskip

{\bf ACKNOWLEDGEMENTS}

\hspace{0.33in} One of the authors (NZ) would like to thank the Arab Fund
for Economic and Social Development for the financial support and the staff
of the I.U.S.T.I., Universit\'{e} de Provence - Marseille for hospitality
during the progress of this work. We thank Professor A.M.Dykhne for fruitful
discussion on the subject.

\newpage

{\Large {\bf Figure Captions}}

\bigskip

{\bf Figure 1\qquad }Predicted temporal evolution (between $1850$ and $2100$%
) of the cases (a) and susceptibles (b) by thousands for $\xi =0.5$ and two
different first infection dates: $T=2$ days and $T=20$ years. The arrows
show the dates of introduction of the first and second vaccination
schedules. \bigskip

{\bf Figure 2\qquad }Predicted temporal evolution (from $1850$ to $2020$) of
the cases a) and susceptibles by thousands for three different values of $%
\xi $: $0.3$, $1$ and $2$. The arrows show the dates of introduction of the
first and second vaccination schedules. \bigskip

{\bf Figure 3\qquad }Predicted temporal evolution (from $1990$ to $2020$) of
the cases (lower curves) and susceptibles (upper curves) by thousands for $%
\xi =0.5$ and two {\bf \ }different ages of administration of the second
vaccination dose: $6$ years (solid curves) and $2$ years (dotted curves).
\bigskip

{\bf Figure 4 \qquad }The incidence normalized to the infected and
susceptibles ($x10^{-7}$per day) as a function of the number of susceptibles
by thousands for $\xi =0.5$ in four different situations of the propagation
of the disease: a) before the first epidemic peak, b) after the first
epidemic peak, c) before an epidemic peak choosed in the equilibrium and d)
after this epidemic peak. The solid lines show linear fits. \bigskip

\newpage

{\Large {\bf Table Captions}}

\bigskip {\bf Table 1\qquad }: Comparison of the simulation for $\xi =0.5$
and a vaccination at $6$ years with the data of Oran (Algeria) during the
period ($1994-1997$).

\newpage

\begin{tabular}{lllll}
Year & $1994$ & $1995$ & $1996$ & $1997$ \\ 
Simulation data & $90$ & $773$ & $219$ & $816$ \\ 
Data of Oran [14] & $86$ & $655$ & $179$ & $286$%
\end{tabular}


\newpage

\begin{thebibliography}{99}
\bibitem{Berno}  D.Bernoulli, {\it Essai d'une nouvelle analyse de la
mortalit\'{e} caus\'{e}e par la petite v\'{e}role et des avantages de
l'inoculation pour la pr\'{e}venir}, in M\'{e}moires de Math\'{e}matiques et
de Physique, Acad\'{e}mie Royale des Sciences, Paris, 1760, pp. 1-45.

\bibitem{Hamer}  W.H.Hamer, Lancet, {\bf 1} (1906) 733.

\bibitem{Baile}  N.T.Bailey, {\it The mathematical theory of infectious
disease, }2nd ed. (Hafner, New York) 1975.

\bibitem{Ander}  R.M.Anderson and R.M.May, {\it Infectious diseases of
humans, dynamics and control} (Oxford University Press, Oxford) 1991.

\bibitem{Hethc}  H.W.Hethcote, SIAM Review, {\bf 42} (2000) 599.

\bibitem{Watts}  D.J.Watts, {\it Small worlds: the dynamics of networks
between order and randomnes}, (Princeton University Press, New Jersey) 1999;
D.J.Watts and S.H.Strogatz, Nature, {\bf 393} (1998) 440; S.H.Strogatz,
Nature, {\bf 410} (2001) 268.

\bibitem{Bocca}  N.Boccara, K.Cheong and M.Oram, J.Phys.A: Math.Gen., {\bf 27%
} (1994) 1585.

\bibitem{Zekri}  N.Zekri and J.P.Clerc, to appear in Phys.Rev.E, (2001).

\bibitem{Rober}  M.G.Roberts and M.I.Tobias, Epidemiol.Infect. {\bf 124}
(2000) 279; G.N.Becker and A.Bahrampour, Math. Biosc. {\bf 142} (1997) 63.

\bibitem{Levy}  D.L\'{e}vy-Bruhl, J.Maccario, S.Richardson and N.Gu\'{e}rin,
Bull.Epid\'{e}miol.Hebd. {\bf 29} (1997) 133; N.G.Becker and V.Rouderfer,
Math.Biosc. {\bf 131} (1996) 81.

\bibitem{Geant}  GEANT, CERN software library ; see also N.Zekri, PhD
thesis, Universit\'{e} de Savoie, 1988.

\bibitem{cdc}  www.cdc.gov/publication.htm.

\bibitem{Mlean}  A.R.Mc Lean and R.M.Anderson, Epidemiol. Inf., {\bf 100}
(1988) 419.

\bibitem{Benth}  A.Benthabet, SEMEP-Oran (Algeria), Private communication.
\end{thebibliography}
\end{document}